\documentclass[english]{iopart}
\usepackage{graphicx}
\usepackage{iopams,setstack}
\usepackage{epsfig}
\usepackage{graphics}
\begin{document}
\title{Spin diffusion in Fermi gases}
\author{G.\ M.\ Bruun}
\address{Department of Physics and Astronomy, University of Aarhus, Ny Munkegade, 8000 Aarhus C, Denmark}

\begin{abstract}
We examine spin diffusion in a two-component homogeneous Fermi gas in the normal phase.
 Using a variational approach, analytical results 
are presented for the spin diffusion coefficient and the related  spin relaxation time
as a function of temperature and interaction strength. For low temperatures, 
 strong correlation effects are included through the Landau parameters which we extract from Monte Carlo results. We show that 
the spin diffusion coefficient has a minimum for a temperature somewhat below the Fermi temperature with a value that  
 approaches the quantum limit $\sim\hbar/m$  in the unitarity regime where $m$ is the particle mass. 
We finally derive a value for the low temperature shear viscosity in the normal phase from the Landau parameters.
\end{abstract}
\maketitle
%\tableofcontents{}
\section{Introduction}
Cold atomic gases provide a unique possibility to study  many-body physics  in a controlled way as 
 one can tune  properties such as the strength and sign of the interaction as well as the population in various internal  states. 
 Fermi gases in the unitarity regime characterized by 
a diverging scattering length $|a|\rightarrow\infty$ are intensely investigated. They represent a universal realization 
of a strongly interacting Fermi gas in the sense that their properties are independent of the details of the interaction. 
Thermodynamic as well as microscopic equilibrium quantities have been measured in a series of impressive 
experiments~\cite{ENS,Duke,MIT,Rice, Tokyo,JILA}. 
For high temperatures, one can calculate their properties in a controlled way using the virial expansion~\cite{Ho,Yu,Drummond}. 
For low temperatures, the system is strongly interacting and poses a challenging problem.  Quantum Monte Carlo calculations 
are therefore presently  the most reliable method to extract their equilibrium properties~\cite{Carlson, MC, Burovski}. 

There is increasing focus on the transport properties of these gases. 
As opposed to thermodynamic quantities, strong interactions can change 
transport coefficients by orders of magnitude, making them attractive to study. 
The shear viscosity $\eta$ of a Fermi gas has been extracted from the damping of  collective modes~\cite{JohnThomas1} as well
 as expansion experiments~\cite{JohnThomas2}.   One can 
calculate the viscosity accurately for a Fermi gas in the unitarity limit for high temperatures~\cite{BruunSmith}. 
Furthermore, for  $T\ll T_c$ with $T_c$ the critical temperature for superfluidity, the contribution to the viscosity from phonons
has been calculated for superfluid $^4$He~\cite{Landau} and for a superfluid Fermi gas in the unitarity regime~\cite{Rupak}.
These two results agree if the sound velocity of a Fermi gas is inserted in the expression for  $^4$He
showing that the microscopic origin of suprfluidity is irrelevant~\cite{Ens}. At intermediate temperatures, it is very challenging to calculate $\eta$ 
from a microscopic theory, since many  effects have to be considered including particle self-energies~\cite{BruunSmith2},  pairing, 
and vertex corrections~\cite{Ens,Guo}.  
Using methods from from string theory, it has been shown that the viscosity of a class of strongly interacting many-body systems 
obey the bound $\eta/s>\hbar/4\pi k_B$  where $s$ is the entropy density~\cite{Policastro}, and it has been 
conjectured that this bound holds for all fluids~\cite{Kovtun}. This has made the study of the viscosity of cold atomic Fermi gases in the unitarity limit directly 
relevant to other fields including quark-gluon and high energy physics~\cite{SchaeferTeaney}.

Very recently, measurements of spin transport properties of a two-component Fermi gas have been 
reported~\cite{ZwierleinSpinTransport}. 
Inspired by this, we analyze in this paper the spin diffusion coefficient $D$ of a two-component homogenous Fermi gas in the 
normal phase.
Using a variational approach, we calculate $D$  both for weak and for strong interactions. For high temperatures, 
 our calculations are analogous to the ones leading to the accurate result for the shear viscosity~\cite{BruunSmith}. 
In the low temperature regime where the system is strongly correlated,  we use Fermi liquid theory 
 to calculate $D$. The strong coupling effects are contained in the 
 Landau parameters which are extracted from Monte Carlo calculations. In this way, we expect to 
obtain reliable results for $D$ even in the unitarity limit. It follows from our high and low $T$ results that 
 the spin diffusion coefficient exhibits a minimum for a temperature below $T_F$. 
 In the unitarity limit, we find that the minimum is $D\sim \hbar/m$. Since 
  $D$ in general decreases with increasing coupling strength, this value can be interpreted as the quantum limit
  for how small $D$ can become in atomic gases, when interactions are as strong as quantum mechanics allows. 
  In analogy with the intriguing conjecture of a global minimum bound for $\eta$, 
   it would be very interesting to    compare this quantum minimum for $D$ with experimental results as well as with $D$ for other strongly interacting
 systems.  We also calculate the closely related spin relaxation time, and  briefly discuss how $\eta$
 can be extracted from the Landau parameters.

\section{Spin diffusion in the hydrodynamic regime} 
We consider a gas of fermions of mass m in two internal states which we denote  spin $\sigma=\uparrow,\downarrow$. 
In equilibrium, the  densities of the two components are equal,  $n_\uparrow=n_\downarrow=n/2=k_F^3/6\pi^2$, with $n$
the total density. We shall focus on the non-equilibrium  
situation with a spatially varying magnetization $n_\uparrow({\mathbf r})-n_\downarrow({\mathbf r})$. 
The corresponding  spin current  ${\mathbf{j}}={\mathbf{j}}_\uparrow-{\mathbf{j}}_\downarrow$
 is in the hydrodnamic regime given by Fick's law
\begin{equation}
{\mathbf j}=-D\nabla(n_\uparrow-n_\downarrow)
\label{Fick}
\end{equation}
where $D$ is the spin diffusion coefficient. A main goal of this paper is to calculate $D$ as a function of coupling strength and temperature. 

In a vacuum, the interaction between  atoms with spins $\uparrow$ and $\downarrow$
is  given by the cross section
\begin{equation}
\frac{d\sigma_{\rm sc}^{\uparrow\downarrow}}{d\Omega}=\frac{a^2}{1+p_r^2a^2}.
\label{ScatVac}
\end{equation} 
with $a$ is the $s$-wave scattering length and $p_r$ the relative momentum. 
There are many-body corrections to (\ref{ScatVac}) but they are small for $T\gg T_F$ with $T_F$ the Fermi temperature of the gas,
 even in the unitarity limit~\cite{BruunFewBody,BruunSmith}. 
 For the densities and temperatures of interest, we can ignore the bare interaction between atoms with equal spins. 
For high temperature, the typical scattering momenta scale as 
$\sim\sqrt{mk_BT}$ and it follows from (\ref{ScatVac}) that the gas is weakly interacting irrespective of the value of $a$.

For low $T$,
 the gas is strongly interacting in the unitarity regime $|a|\rightarrow\infty$  and there is presently no quantitatively reliable microscopic theory for calculating its  transport properties.
Assuming the gas is in the normal phase for $T\ll T_F$, we can however use Fermi liquid theory. Combined with 
the relevant Landau parameters extracted from non-perturbative Monte-Carlo calculations, we can in this way 
derive  reliable results for $D$ in the strongly interacting normal phase~\cite{BaymPethick, BaymEbner}. Since 
the gas is superfluid for $T<T_c$ and $T_c$ is predicted to 
be a sizable fraction of $T_F$ in the unitarity limit~\cite{Burovski}, it is not obvious that one can fulfill the criterion $T_c<T\ll T_F$ for a
normal Fermi liquid.   However, we will show that the minimum of $D$ seems to be located for $T>T_c$ where pairing is not relevant. 
Also, one can in fact quench the superfluid order for $T<T_c$ by rotating the gas~\cite{Bausmerth}.

\subsection{The Landau-Boltzmann equation}
We proceed using kinetic theory for high $T$ and Fermi liquid theory for low $T$ to describe the spin
dynamics of the gas.  In the hydrodynamic regime, the typical length scale of the dynamics is much longer than the mean free path. 
The non-equilibrium distribution function is then close to a local hydrodynamic form, i.e.\ 
$f_\sigma({\mathbf{r}},{\mathbf{p}})\simeq1/\{\exp[\epsilon_{\mathbf p}-\mu_\sigma({\mathbf{r}})]+1\}$ with $\epsilon_{\mathbf p}$ 
the quasiparticle energy and $\mu_\uparrow({\mathbf{r}})$ and  $\mu_\downarrow({\mathbf{r}})$ spatially varying chemical potentials  
 corresponding to the magnetization  $n_\uparrow({\mathbf r})-n_\downarrow({\mathbf r})$.
  Plugging the local equilibrium distributions  into the left side of the steady state linearized 
Landau-Boltzmann equation for the two spin components and taking the difference yields~\cite{BaymPethick,SmithHojgaard}
\begin{equation}
\beta{\mathbf{v_p}}\cdot\nabla(\mu_\uparrow-\mu_\downarrow)=-\frac{I_\uparrow-I_\downarrow}{f^0(1-f^0)}
\label{BEdiff}
\end{equation} 
with ${\mathbf{v_p}}=\nabla_{\mathbf p}\epsilon_p$  the velocity and $f^0$ the distribution function in equilibrium. 
The difference in the linearized collision integrals is
\begin{equation}
\fl
I_\uparrow-I_\downarrow
=\int \frac{d^3 p_2}{(2\pi)^3} d\Omega\frac{|{\mathbf{p}}-{\mathbf{p}}_2|}{m^*}%\left[
\frac{d\sigma_{\rm sc}^{\uparrow\downarrow}}{d\Omega}(\Psi-\Psi_2-\Psi_3+\Psi_4)%+\frac 1 2 \frac{d\sigma_{\rm sc}^{\uparrow\uparrow}}{d\Omega}(\Psi+\Psi_2-\Psi_3 \Psi_4)\right]
f^0f_2^0(1-f^0_3)(1-f_4^0)
\label{Icoll}
\end{equation} 
with $m^*$ the effective mass. We have written 
$\delta f_\uparrow({\mathbf{r}},{\mathbf{p}})-\delta f_\downarrow({\mathbf{r}},{\mathbf{p}})= f^0({\mathbf{r}},{\mathbf{p}})[1-f^0({\mathbf{r}},{\mathbf{p}})]\Psi_\sigma({\mathbf{r}},{\mathbf{p}})$ with $\delta f_\sigma$ the deviation of the distribution function away from local equilibrium.  
 The differential cross section for the scattering
between particles with opposite spins is $d\sigma_{\rm sc}^{\uparrow\downarrow}/d\Omega$ with $\Omega$  the solid 
angle between the outgoing and ingoing relative momenta, ${\mathbf p}_r'=({\mathbf p}_4-{\mathbf p}_3)/2$ and 
${\mathbf p}_r=({\mathbf p}_2-{\mathbf p})/2$. There is also a term in (\ref{Icoll}) coming from the 
induced interaction between parallel spin quasiparticles which 
we have not written explicitly, since it does not contribute to the variational  results for $D$ given below. 
In the following, we take the magnetization to vary in the $z$-direction. The spin current is then given by 
\begin{equation}
j_z=\int \frac{d^3 p}{(2\pi)^3}(\delta f_\uparrow-\delta f_\downarrow)v_z.
\label{current}
\end{equation}

\subsection{A variational expression for $D$}
The Landau-Boltzmann equation (\ref{BEdiff}) can be written as $\kappa=H[\Phi]$ with $H=(I_\uparrow-I_\downarrow)/f^0(1-f^0)$
and $\kappa=-\beta v_z\partial_z(\mu_\uparrow-\mu_\downarrow)$. Likewise, the  spin current (\ref{current})
can be written as $-\langle\kappa\Psi\rangle k_BT/\partial_z(\mu_\uparrow-\mu_\downarrow)$
with the definition $\langle A\rangle=\int d^3p f^0(1-f^0)A/(2\pi)^3$. Using this, 
 one can in analogy with the case of charge current~\cite{SmithHojgaard, MBS}  derive a variational bound for the spin current given by
\begin{equation}
j_z\ge-\frac{k_BT}{\partial_z(\mu_\uparrow-\mu_\downarrow)}\frac{\langle U\kappa\rangle^2}{\langle UH[U]\rangle}.
\label{Varcurrent1}
\end{equation}
Here, $U$ is a trial function for the deviation function $\Psi$. 
By expanding $U$ in polynomials,   
 it has been shown that using the driving term in the Landau-Boltzmann equation as an ansatz, i.e.   $U\propto\kappa$,
  yields very accurate results for the viscosity $\eta$~\cite{BruunSmith2}.
We will therefore use the ansatz $U=p_z\propto\kappa$ appropriate for spin diffusion in the following. This  gives
 \begin{equation}
D=\frac\beta{{m^*}^2}\frac{1}{\chi_s}
\frac{\langle p_z^2\rangle^2}{\langle p_zH[p_z]\rangle}
\label{VarD}
\end{equation}
 as our variational expression for the spin diffusion coefficient with $\chi_s=\partial (n_\uparrow-n_\downarrow)/\partial(\mu_\uparrow-\mu_\downarrow)$
 the spin susceptibility.  Equation (\ref{VarD}) serves as the starting point for  our results concerning  spin diffusion.
It is in fact the use of the ansatz $U=p_z$ combined with  momentum conservation, which gives that 
 it is only the scattering between opposite spins that  enters in the variational expression for $D$.

\section{High temperature limit}
We now calculate $D$ in the classical limit $T\gg T_F$. 
 Using  $f^0=\exp[-\beta(p^2/2m-\mu)]\ll 1$ with $\mu$ the chemical potential in equilibrium,
   the integrals in (\ref{VarD}) can performed and we obtain 
 \begin{equation}
 \fl
D=\frac{3\sqrt{\pi}}{8}\frac{\sqrt{k_BT}}{\sqrt mn\bar\sigma_{\rm sc}^{\uparrow\downarrow}}
=\frac{9\pi^{3/2}\hbar}{32\sqrt 2 m}\times\left\{\begin{array}{ll}
 \frac{1}{(k_Fa)^2}\left(\frac{T}{T_F}\right)^{1/2} &{\rm{ for }}\quad T\ll T_a\;{\rm{ (weak\; coupling)}}\\ 
\left(\frac{T}{T_F}\right)^{3/2} &{\rm{ for }}\quad T\gg T_a\; {\rm{ (unitarity\; limit)}} 
\end{array}
\right. .
\label{DHighT}
\end{equation}
  Here, 
\begin{equation}
\bar\sigma_{\rm sc}^{\uparrow\downarrow}=4\pi a^2\int_0^\infty dx \frac{x^5e^{-x^2}}{1+x^2T/T_a}
\end{equation}
 with $k_BT_a=\hbar^2/ma^2$ is the thermal average of the scattering cross section weighted with $p^3$. We have 
 $\bar\sigma^{\uparrow\downarrow}_{\rm sc}=4\pi a^2$ and 
 $\bar\sigma^{\uparrow\downarrow}_{\rm sc}=2\pi\hbar^2/mk_BT$ in the weak coupling and unitarity limits respectively. 
To derive (\ref{DHighT}), we have used 
$\chi_s=n/2k_BT$ for the spin susceptibility in the classical limit.   The temperature dependence in (\ref{DHighT}) can be understood as follows:
  The spin diffusion coefficient scales as $D\sim l_{\rm mf} v$ with $l_{\rm mf}=2/n\sigma_{\rm sc}$ the mean free path and  $v$ a typical velocity. 
  In the classical regime,  $l_{\rm mf}\sim 1/na^2$ for weak coupling and $l_{\rm mf}\sim mk_BT/\hbar^2n$  in the unitarity limit. Since
  $v\sim \sqrt{k_BT/m}$, we recover the $T^{1/2}$ and $T^{3/2}$ scaling in (\ref{DHighT}). 
  The temperature scaling is analogous to what is found for the shear viscosity~\cite{BruunSmith}. 
    The result (\ref{DHighT}) for a gas in the classical limit $T\gg T_F$ has also recently been derived in Ref.~\cite{ZwierleinSpinTransport}.

\section{Low temperature limit}
For $T\ll T_F$, the quasiparticle scattering takes place around the Fermi surface. The collision integral 
can then be reduced to 
  \begin{equation}
\langle p_zH[p_z]\rangle=\frac{{m^*}^2k_F^3}{18\pi^2}(k_BT)^3I_{\rm angle}
\label{Icoll3}
\end{equation} 
where 
  \begin{equation}
I_{\rm angle}=\int_0^\pi d\theta\frac{\sin \theta(1-\cos\theta)}{\cos(\theta/2)}\int_0^{2\pi} d\phi \frac{d\sigma_{\rm sc}^{\uparrow\downarrow}}{d\Omega}
(1-\cos\phi)
\label{Iangle}
\end{equation} 
is the cross section averaged over the Fermi surface. Here,  $\theta$ is the angle between the incoming scattering momenta ${\mathbf p}$
and ${\mathbf p}_2$ and  $\phi$ is the angle between the relative momenta ${\mathbf p}_r$ and ${\mathbf p}_r'$. 
The cross section at the Fermi surface can be written as
 \begin{equation}
\frac{d\sigma^{\uparrow\downarrow}(\theta,\phi)}{d\Omega}=\frac{\pi^2}{16k_F^2} 
[N(0){\mathcal{T}}^{\uparrow\downarrow}(\theta,\phi)]^2
\label{crosssecupdown}
 \end{equation}
where  ${\mathcal{T}}^{\uparrow\downarrow}$ is the scattering matrix for opposite spins and 
$N(0)=m^*k_F/\pi^2$ is the density of states at the Fermi level. 
The scattering matrix  can be parametrized in terms 
of the Landau parameters $F_l^i$ as~\cite{BaymPethick}
 \begin{equation}
N(0){\mathcal{T}}^{\uparrow\downarrow}(\theta,\phi)=\frac{1}{2}\sum_l[A_l^s-3A_l^a+(A_l^s+A_l^a)\cos\phi]P_l(\cos\theta)
\label{tupdown}
 \end{equation}
 where $P_l(x)$ are the Legendre polynomials and 
\begin{equation}
A_l^i=\frac{F_l^i}{1+F_l^i/(2l+1)}.
\label{A}
\end{equation}
In (\ref{A}), the symmetry of the triplet part of the scattering is taken into account by the $\cos\phi$ factor~\cite{Dy}.
 We now assume that it is sufficient to use the $l=0$ Landau parameters. 
 Using (\ref{crosssecupdown})-(\ref{tupdown}) in (\ref{Iangle}) and solving the resultant trigonometric integrals, 
 yields  when plugged into (\ref{VarD}) 
\begin{eqnarray}
%\fl
D&=&\frac{24\hbar}{\pi^3m}\frac{1+F_0^a}{C_1^2+C_3^2/2-C_1C_3}\left(\frac{T_F}{T}\right)^2\nonumber\\
&=&\frac{3\hbar}{8\pi m}\left(\frac{T_F}{T}\right)^2\times
\left\{\begin{array}{ll}
\frac{1}{(k_Fa)^2}&{\rm{for }}\quad k_Fa\ll 1\;{\rm{ (weak\; coupling)}}\\ 
2.6 &{\rm{ for }}\quad k_Fa\gg 1 \;{\rm{ (unitarity\; limit)}} 
\end{array}
\right. .
\label{DlowTLandau}
\end{eqnarray}
We have defined $C_1=A_0^s-3A_0^a$ and $C_3=A_0^s+A_0^a$ and used $\langle p_z^2\rangle=k_BTm^*k_F^3/6\pi^2$,   and 
$\chi_s=N(0)/2(1+F_0^a)$ for $T/T_F\ll 1$ to derive (\ref{DlowTLandau}).  Also, $m^*=m$ when $F_1^s=0$.
For $k_F|a|\ll 1$, one has $F_0^s=2k_Fa/\pi$ and $F_0^a=-2k_Fa/\pi$ which gives the 
weak coupling result in  (\ref{DlowTLandau}).
The spin diffusion coefficient increases as $T^{-2}$ for low T. This is the usual effect of Fermi blocking of the 
scattering which enables the quasiparticles to travel further thereby increasing the spin current.

 To obtain the numerical value in (\ref{DlowTLandau})  for $k_F|a|\gg 1$, 
we have used values for the Landau parameters 
extracted from Monte Carlo calculations for a strongly interacting two-component Fermi gas:   
 $F_0^s\simeq-0.44$ and $F_0^a\simeq2$~\cite{Carlson,Stringari} giving $A_0^s\simeq-0.79$ and $A_0^a\simeq0.7$. 
Since $A_0^s+A_0^a\simeq-0.1$ almost fulfills the sum rule $\sum_l(A_l^s+A_l^a)=0$, it seems that 
it is reasonable only to use the $l=0$ Landau parameters. 
Note that the interaction between opposite spins  (\ref{tupdown}) is attractive for these Landau parameters:
$N(0){\mathcal{T}}^{\uparrow\downarrow}\simeq-1.4-0.06\cos\phi$. This reflects the tendency for pairing in the gas 
which suppresses the spin susceptibility by the factor $1/(1+F_0^a)\simeq 1/3$. Since it is the chemical 
potentials and not the magnetization $n_\uparrow-n_\downarrow$
 which drive the Landau-Boltzmann equation (\ref{BEdiff}), the spin diffusion coefficient is increased by the same factor. 

\section{Minimum of the spin diffusion coefficient}
\begin{figure}[t]
\begin{center}
\leavevmode
\begin{minipage}{.49\columnwidth}
\includegraphics[clip=true,height=0.6\columnwidth,width=1\columnwidth]{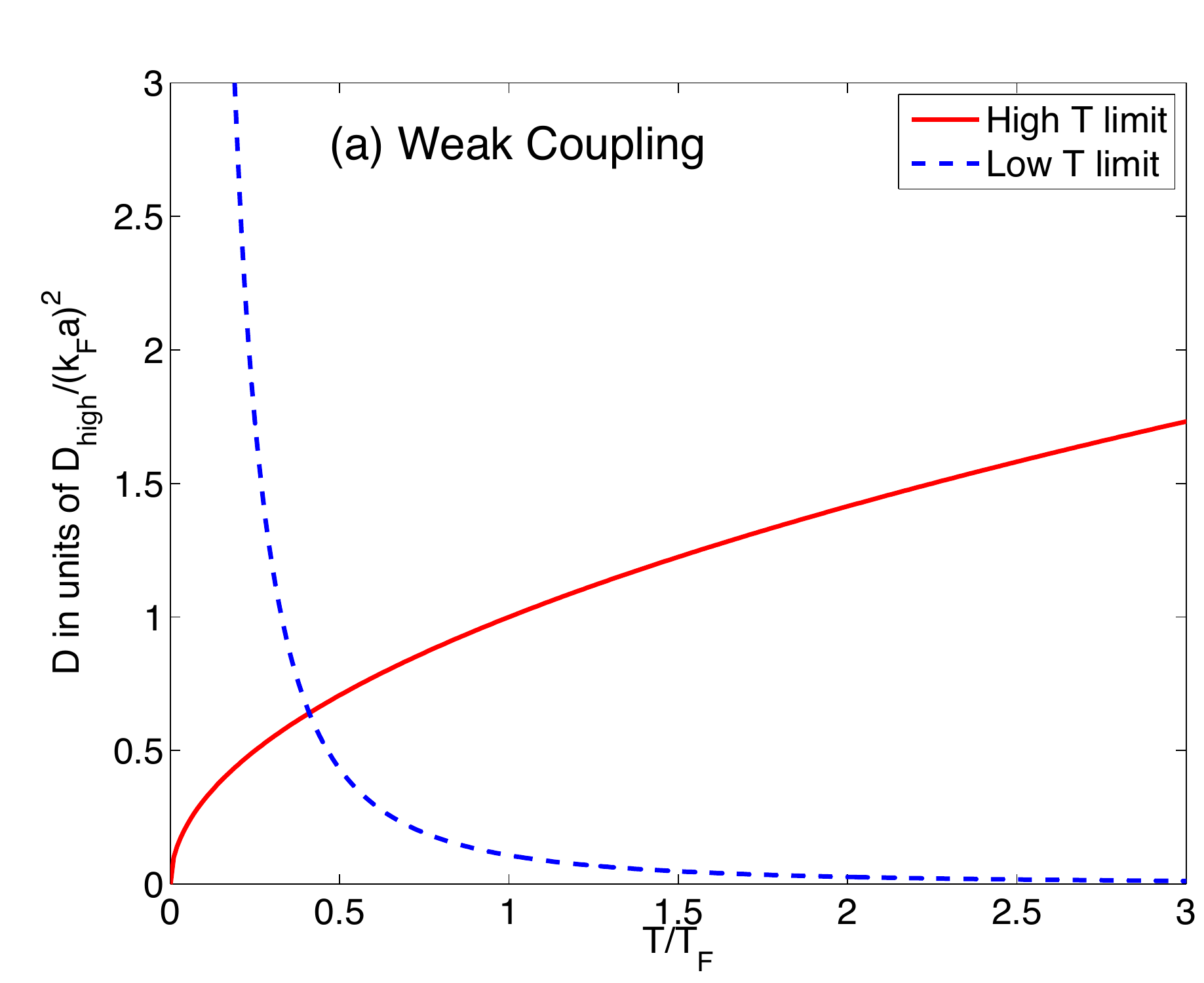}
\end{minipage}
\begin{minipage}{.49\columnwidth}
\includegraphics[clip=true,height=0.6\columnwidth,width=1\columnwidth]{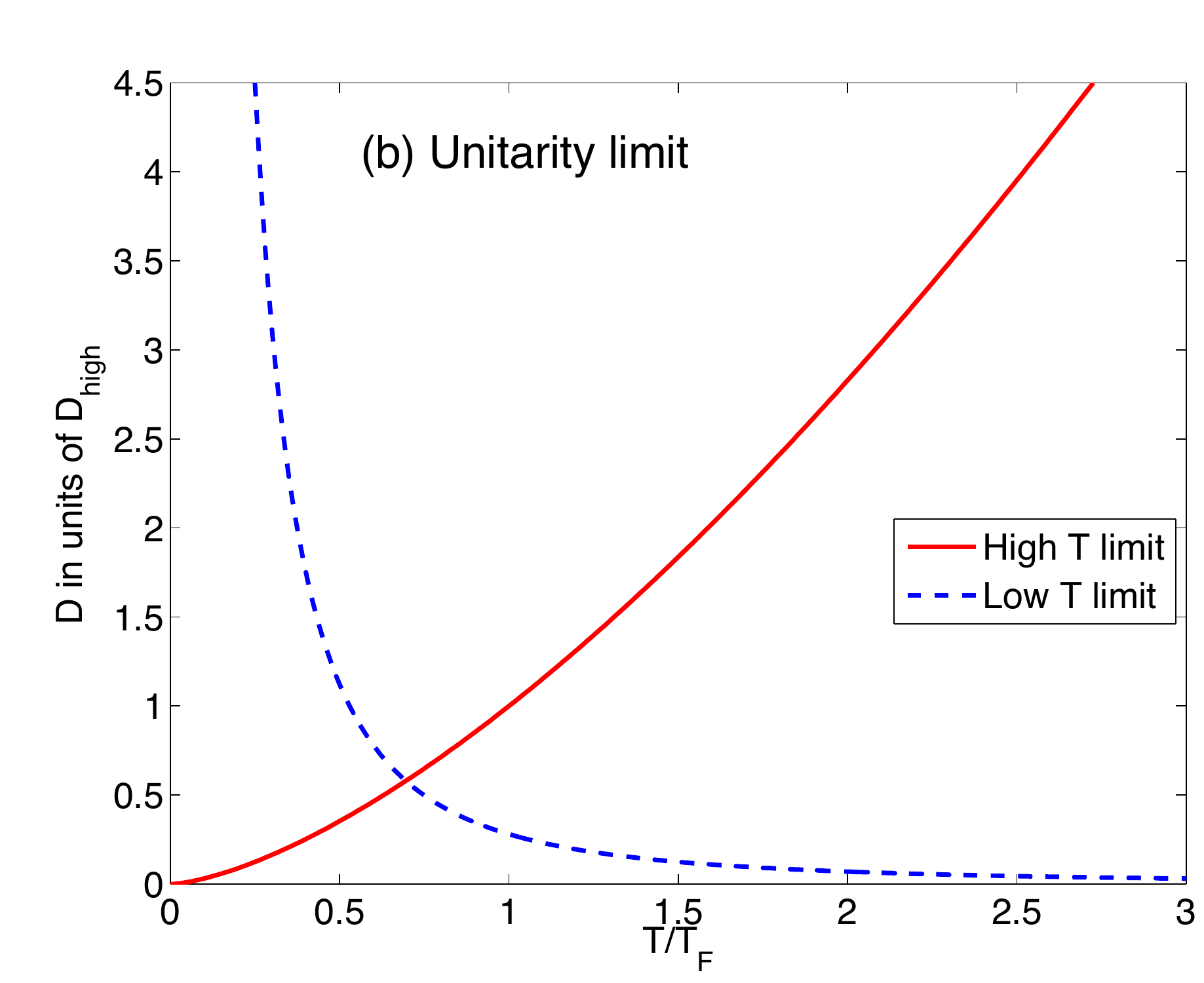}
\end{minipage}
\caption{The spin diffusion coefficient as a function of $T$ in the (a) weak and (b) strong coupling limits.}
\label{DvsT}
\end{center}
\end{figure}

\begin{figure}[t]
\begin{center}
\leavevmode
\begin{minipage}{.49\columnwidth}
\includegraphics[clip=true,height=0.6\columnwidth,width=1\columnwidth]{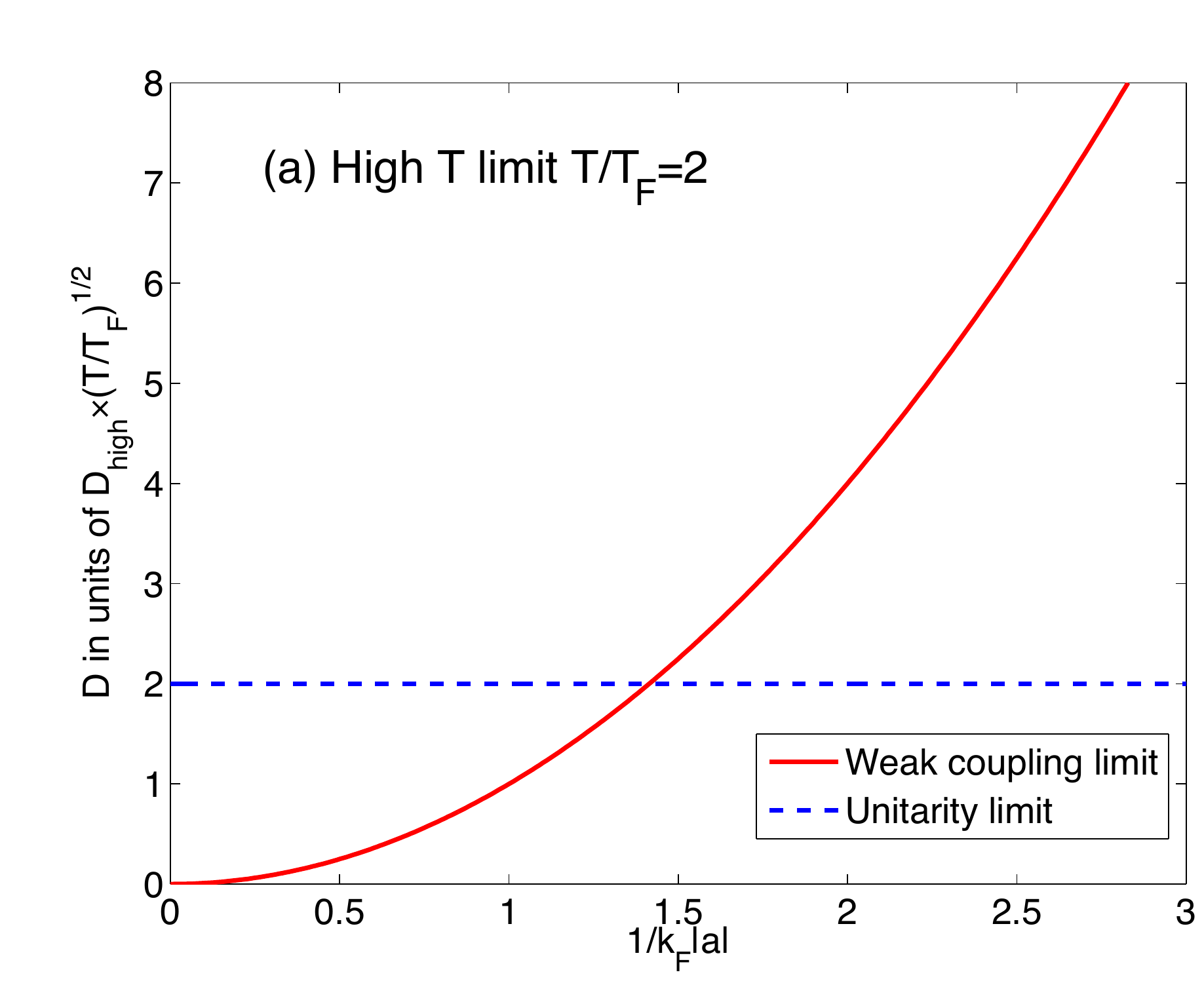}
\end{minipage}
\begin{minipage}{.49\columnwidth}
\includegraphics[clip=true,height=0.6\columnwidth,width=1\columnwidth]{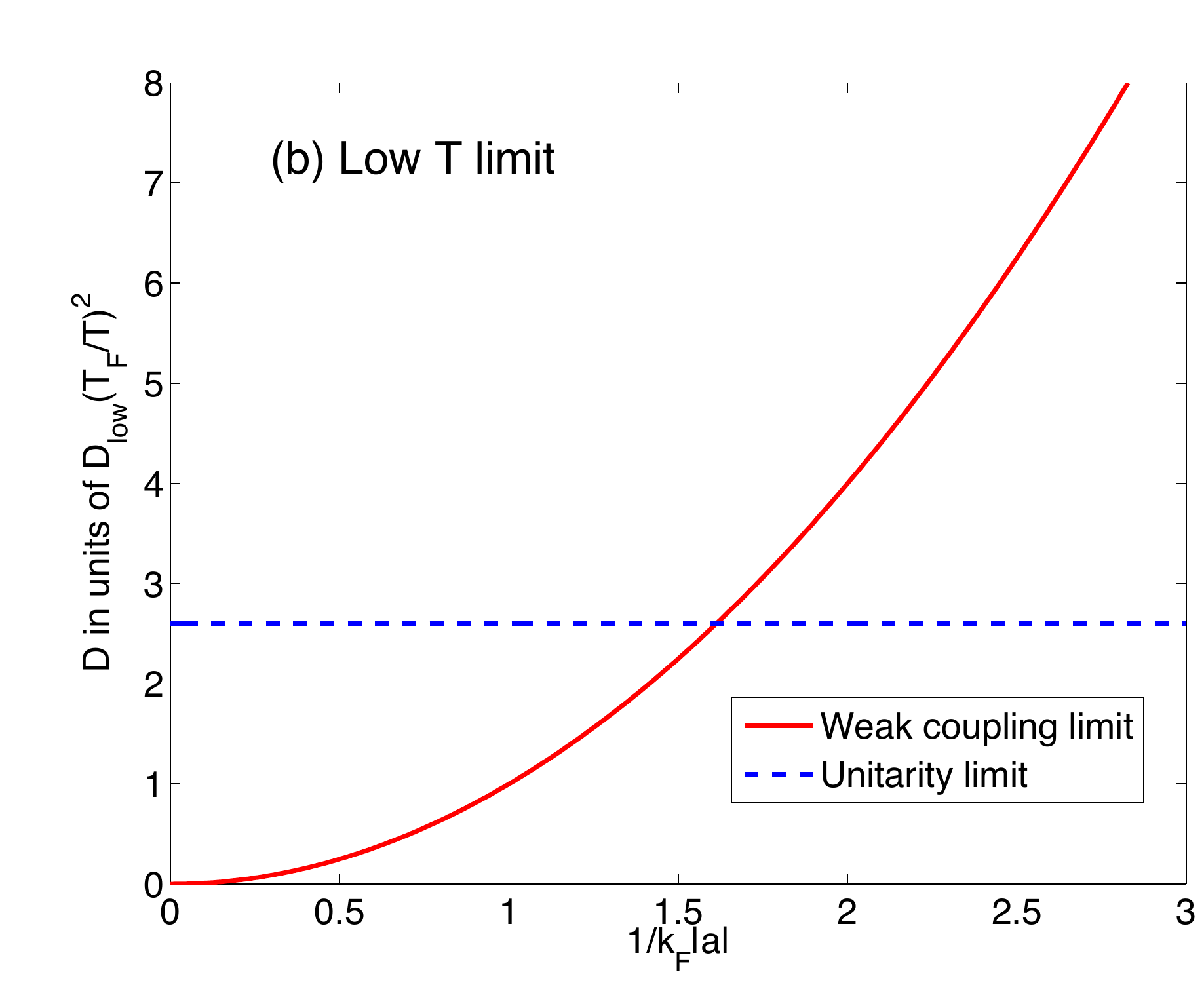}
\end{minipage}
\caption{The spin diffusion coefficient as a function of $1/k_F|a|$ in the (a) high $T$ and (b) low $T$ 
limits.}
\label{Dvsinvkfa}
\end{center}
\end{figure}

In Figs.\ \ref{DvsT}-\ref{Dvsinvkfa}, we summarize our results by plotting in 
$D$ as a function of $T$ and $1/k_F|a|$. In Fig.\ \ref{DvsT} (a), we plot $D$ 
in units of $D_{\rm high}/(k_Fa)^2$ with $D_{\rm high}=9\pi^{3/2}\hbar/32\sqrt 2 m$ as a function of 
$T$ in the weak coupling regime.  
The low and high $T$ curves are given by the weak coupling limits of (\ref{DHighT}) and (\ref{DlowTLandau}).
 Likewise, we plot in Fig.\ \ref{DvsT} (b) $D(T)/D_{\rm high}$ in the unitarity limit as given by (\ref{DHighT}) and (\ref{DlowTLandau}).
Since $D(T)$ will interpolate between these two limits, these results show that $D(T)$ will exhibit a minimum.
Figure \ref{DvsT} indicates that in the unitarity limit,  the minimum is for $T$ somewhat below $T_F$ but above $T_c\sim0.2T_F$
where pairing effects have to be included. 
 The scale of the minimum value is $D_{\rm min}\sim 9\pi^{3/2}\hbar/32\sqrt 2 m\simeq 1.1\hbar/m$.
Since $D$ decreases with increasing coupling strength, this result 
 can be regarded as the quantum limit of $D$ when the interactions are as strong as quantum mechanics allows. 
%%%%%%%%%%%%%%%%%%%%%%%%%%%%%%%%%%%%%%%%%%%%%%%%
%%%%%%%%%%%%%%%%%%%%%%%%%%%%%%%%%%%%%%%%%%%%%%%%

In Fig.\ \ref{Dvsinvkfa}, we plot $D$ as a function of $1/k_F|a|$ in the high temperature limit (a) and in the low temperature limit (b) 
using the units $D_{\rm high}\times (T/T_F)^{1/2}$ and $3\hbar/8\pi m\times(T_F/T)^2=D_{\rm low}\times(T_F/T)^2$ 
respectively. The weak and strong coupling results are again obtained from (\ref{DHighT}) and (\ref{DlowTLandau}), and $D(1/k_F|a|)$
 will  interpolate between these two limits. We see that $D$ decreases with increasing coupling strength $k_F|a|$ and that it 
 eventually flattens 
 out in the limit $k_F|a|\rightarrow\infty$. 
 %%%%%%%%%%%%%%%%%%%%%%%%%%%%%%%%%%%%%%%%%%%%%%%%
%%%%%%%%%%%%%%%%%%%%%%%%%%%%%%%%%%%%%%%%%%%%%%%%

\section{Spin relaxation time}
\begin{figure}[t]
\begin{center}
\leavevmode
\begin{minipage}{.49\columnwidth}
\includegraphics[clip=true,height=0.6\columnwidth,width=1\columnwidth]{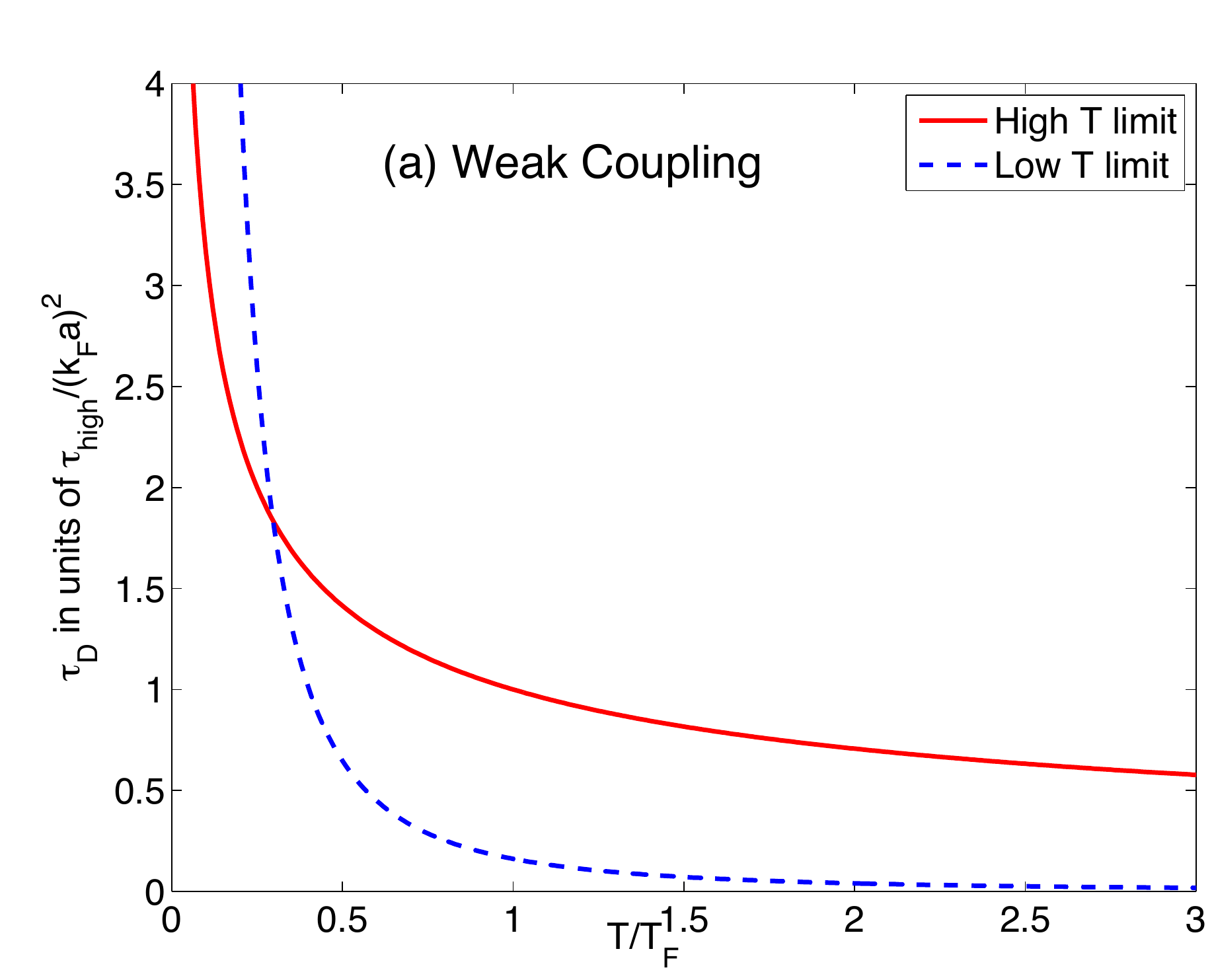}
\end{minipage}
\begin{minipage}{.49\columnwidth}
\includegraphics[clip=true,height=0.6\columnwidth,width=1\columnwidth]{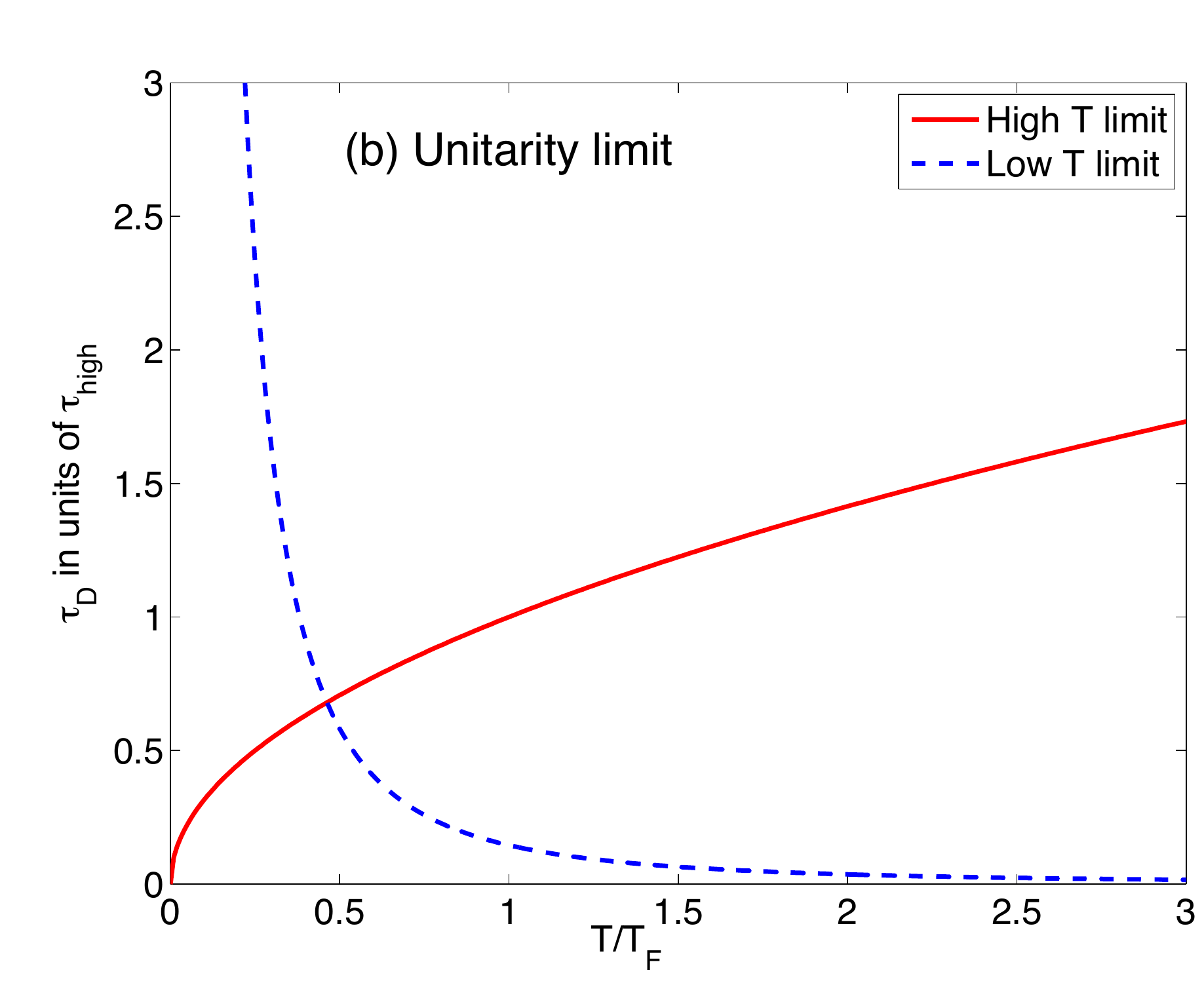}
\end{minipage}
\caption{The spin relaxation time as a function of $T$ in the (a) weak and (b) strong coupling limits.}
\label{TauvsT}
\end{center}
\end{figure}

From our results, we now derive 
 an expression for the spin relaxation time which gives the typical time between scattering events for spin dynamics. 
 A suitable definition of $\tau_D$ can be obtained from the relaxation time approximation, i.e.\ by assuming 
 $I_\uparrow-I_\downarrow=-(\delta f_\uparrow-\delta f_\downarrow)/\tau_D$. Using this, 
  the linearized Landau-Boltzmann equation (\ref{BEdiff})
is easily solved for $\delta f_\uparrow-\delta f_\downarrow$. Plugging into (\ref{current}) then gives the usual expression 
$D=\frac 1 3(1+F_0^a)v_F^2\tau_D$ for $T\ll T_F$ with $v_F$ the Fermi velocity. 
We \emph{define} the spin diffusion time $\tau_D$ by comparing this to  (\ref{DlowTLandau}) which gives 
\begin{equation}
\fl
\tau_D=\frac{9\hbar}{16\pi k_BT_F}\left(\frac{T_F}{T}\right)^2\times\left\{\begin{array}{ll}
 \frac{1}{(k_Fa)^2}&{\rm{ for }}\quad k_Fa\ll 1\; {\rm{ (weak\; coupling)}}\\ 
0.9 &{\rm{ for }}\quad k_Fa\gg 1\; {\rm{ (unitarity\; limit)} }
\end{array}
\right. .
\label{TauDLowT}
\end{equation}
for $T\ll T_F$. Likewise,  the relaxation time approximation yields $D=k_BT\tau_D/m$ for $T\gg T_F$.  When comparing to (\ref{DHighT}) this gives
\begin{equation}
\fl
\tau_D=\frac{9\pi^{3/2}\hbar}{32\sqrt 2 k_BT_F}\times\left\{\begin{array}{ll}
 \frac{1}{(k_Fa)^2}\left(\frac{T}{T_F}\right)^{-1/2} &{\rm{ for }}\quad T\ll T_a\;{\rm{ (weak\; coupling)}}\\ 
\left(\frac{T}{T_F}\right)^{1/2} &{\rm{ for }}\quad\; T\gg T_a\; {\rm{ (unitarity\; limit)}} 
\end{array}
\right. 
\label{TauDHighT}
\end{equation}
for $T\gg T_F$. In Fig.\ (\ref{TauvsT}) we plot the high and low $T$ limits spin relaxation time as given by (\ref{TauDLowT}) and (\ref{TauDHighT}).
%%%%%%%%%%%%%%%%%%%%%%%%%%%%%%%%%%%%%%%%%%%%%%%%
%%%%%%%%%%%%%%%%%%%%%%%%%%%%%%%%%%%%%%%%%%%%%%%%
We have used the units $\tau_{\rm high}/(k_Fa)^2$ for the weak coupling case and $\tau_{\rm high}$ in the unitarity limit with 
$\tau_{\rm high}=9\pi^{3/2}\hbar/32\sqrt 2 k_BT_F$. Again, we see that the spin relaxation time exhibits a minimum for $T<T_F$ 
in the unitarity limit. However, $\tau_D$ increases monotonically with decreasing $T$ in the weak coupling regime. 

The spin relaxation time is useful to  estimate the nature of the spin dynamics in a particular trapped atomic gas experiment:
When $\omega\tau_D\ll 1$ with $\omega$ the relevant trapping frequency, the spin dynamics is hydrodynamic whereas it is collisionless for 
$\omega\tau_D\gg 1$.

\section{Viscosity for low T}
For completeness, we briefly outline how one can obtain the viscosity $\eta$ and the viscous relaxation time $\tau_\eta$ for $T/T_F\ll 1$ for 
a normal gas from the Landau parameters. The variational approach for calculating $\eta$ is explained in Ref.~\cite{MBS}. 
For $T/T_F\ll 1$, the viscosity is determined by the angular average of the cross section over  the Fermi surface as in (\ref{Icoll3})-(\ref{Iangle}). 
For the viscosity however, it is the full scattering cross section 
$d\sigma_{\rm sc}^{\uparrow\downarrow}/d\Omega+\frac 1 2 d\sigma_{\rm sc}^{\uparrow\uparrow}/d\Omega$ which enters. 
Here $d\sigma_{\rm sc}^{\uparrow\uparrow}/d\Omega$ describes the induced interaction between parallel spins which cannot be ignored 
in general, even though the bare interaction (\ref{ScatVac}) is only between opposite spins. 
One has
in analogy with (\ref{crosssecupdown}) 
 \begin{equation}
\frac{d\sigma^{\uparrow\uparrow}(\theta,\phi)}{d\Omega}=\frac{\pi^2}{16k_F^2} 
[N(0){\mathcal{T}}^{\uparrow\uparrow}(\theta,\phi)]^2
\label{crosssecupup}
 \end{equation}
 with
  \begin{equation}
N(0){\mathcal{T}}^{\uparrow\uparrow}(\theta,\phi)=\sum_l(A_l^s+A_l^a)P_l(\cos\phi)\cos\phi. 
\label{tupup}
 \end{equation}
 Taking only the $l=0$ Landau parameters,  we get 
 $N(0){\mathcal{T}}^{\uparrow\uparrow}(\theta,\phi)=(A_0^s+A_0^a)\cos\phi=-0.12\cos\phi $. This means
 that the  induced interaction between parallel spins is attractive but much weaker than the interaction between different spins. Following 
 steps analogous to the spin diffusion case described above, we obtain
  \begin{equation}
\eta=\frac{24}{\pi^3}\frac{T_F^2}{T^2}\frac{1}{C_1^2+\frac{3}{4}C_3^2}n\hbar=0.1\left(\frac{T_F}{T}\right)^2n\hbar.
\label{etafinal}
\end{equation}
for $T\ll T_F$. Using the relaxation time result, $\eta=n\frac{k_F^2}{m^*}\tau_\eta/5$ to define the viscous relaxation time $\tau_\eta$, we get 
\begin{equation}
\tau_\eta=\frac{60}{\pi^3}\frac{T_F^2}{T^2}\frac{1}{C_1^2+\frac{3}{4}C_3^2}\frac{\hbar}{k_BT_F}=0.2\left(\frac{T_F}{T}\right)^2\frac{\hbar}{k_BT_F}.
\label{taueta}
\end{equation}
Comparing (\ref{taueta}) with (\ref{TauDLowT}), we see that the two relaxation times are qualitatively the same 
 as expected. They therefore yield the same prediction for the cross-over between hydrodynamic and 
 collisionless behavior.

\section{Conclusions}
Using a variational approach, we analyzed 
the spin diffusion coefficient and the spin relaxation time for a two-component homogeneous Fermi gas in the hydrodynamic limit.  
We derived analytical results in the high and the low temperature regimes including strong coupling effects through 
Landau parameters extracted from 
Monte Carlo calculations. Our results indicate that the
spin diffusion coefficient exhibits a minimum for a temperature below $T_F$  but above $T_c$, with a  
value which  scales as $\sim\hbar/m$ in the unitarity regime. 
It would be very interesting to compare this result with the value of the spin diffusion 
coefficient in other strongly interacting systems.  Also, one should analyze the effects of pairing on spin diffusion 
in atomic gases. New experimental insight is particularly  relevant 
 due to the lack of a controllable method for calculating the minimum value of $D$ quantitatively
 in the strong coupling limit. 
We finally provided expressions for the shear viscosity in terms of the Landau parameters.

\ack
We acknowledge M.\ Zwierlein for several discussions and for providing his experimental as well as theoretical results.  
We are grateful to C.\ J.\ Pethick for useful discussions.

\end{document}